\documentstyle[12pt]{article}
\begin{document}
\title{{\bf ASPECTS OF QUANTUM COSMOLOGY}
\thanks{Alberta-Thy-14-95, gr-qc/9507025, to be published in {\em Proceedings
of the International School of Astrophysics ``D. Chalonge,'' 4th Course:
String Gravity and Physics at the Planck Energy Scale, Erice, Sicily, 8-19
September 1995}, edited by N. Sanchez (Kluwer, Dordrecht, 1996)}}
\author{
Don N. Page
\thanks{Internet address:
don@phys.ualberta.ca}
\\
CIAR Cosmology Program, Institute for Theoretical Physics\\
Department of Physics, University of Alberta\\
Edmonton, Alberta, Canada T6G 2J1
}
\date{(1995 July 10)}

\maketitle
\large
\begin{abstract}
\baselineskip 14.7 pt

 Quantum mechanics may be formulated as {\it Sensible Quantum Mechanics} (SQM)
so that it contains nothing probabilistic, except, in a certain frequency
sense, conscious perceptions.  Sets of these perceptions can be
deterministically realized with measures given by expectation values of
positive-operator-valued {\it awareness operators} in a quantum state of the
universe which never jumps or collapses.  Ratios of the measures for these sets
of perceptions can be interpreted as frequency-type probabilities for many
actually existing sets rather than as propensities for potentialities to be
actualized, so there is nothing indeterministic in SQM.  These frequency-type
probabilities generally cannot be given by the ordinary quantum
``probabilities'' for a single set of alternatives.  {\it Probabilism}, or
ascribing probabilities to  unconscious aspects of the world, may be seen to be
an {\it aesthemamorphic myth}.

 No fundamental correlation or equivalence is postulated between different
perceptions (each being the entirety of a single conscious experience and thus
not in direct contact with any other), so SQM, a variant of Everett's
``many-worlds'' framework, is a ``many-perceptions'' framework but not a
``many-minds'' framework.  Different detailed SQM theories may be tested
against experienced perceptions by the {\it typicalities} (defined herein) they
predict for these perceptions.  One may adopt the {\it Conditional Aesthemic
Principle}:  among the set of all conscious perceptions, our perceptions are
likely to be typical.
\\
\\
\end{abstract}
\normalsize
\baselineskip 14.7 pt

\section{Basics of Canonical Quantum Cosmology}

\hspace{.25in}Quantum cosmology is quantum theory applied to the whole
universe.  At first sight, this may seem like a strange thing to study, since
quantum theory is supposed to apply to the very small, whereas the universe is
very large.  However, there are at least three motivations for studying quantum
cosmology:

(1) Because it's there.  Quantum theory certainly seems to apply to parts of
the universe (e.g., microscopic systems), and for those parts, the indications
are that it is more basic or fundamental than classical theory, which appears
to arise as some approximation for certain macroscopic systems.  Although we do
not know definitely that quantum theory applies to the entire universe, that
certainly looks like the simplest possibility consistent with our present
knowledge, so it behooves us to investigate the implications if it does.

(2) Our present classical theory of gravity, Einstein's general relativity,
contains within itself the seeds for its own destruction in predicting (given
certain observations about the present universe and certain reasonable
assumptions about the behavior of matter at high densities) that the universe
was once so small and highly curved that classical theory should not have been
valid.  In other words, we believe the universe was once so small that indeed
quantum theory should have been needed to understand it.

(3) Present properties of the universe can be described but not explained by
classical theory:  (a) the flatness of the universe, meaning its large volume
and number of particles; (b) the approximate isotropy and homogeneity of the
large-scale universe; (c) the particular form of the inhomogeneous structure of
the universe on smaller scales; and (d) the thermodynamic arrow of time.

Since the dominant interaction on the largest scales of the universe is
gravity, quantum cosmology inevitably involves quantum gravity in a fundamental
way.  Unfortunately, we do not yet have a complete consistent theory of quantum
gravity.  Superstring theory seems to be the best current candidate for
becoming such a theory, but it is not yet well understood, particularly at the
nonperturbative level.  Part of the problem of quantum gravity is to get a
theory which remains calculable (e.g., can be rendered finite) when one
includes arbitrarily high energy fluctuations, and for this problem superstring
theory does seem to be remarkably successful, at least at the perturbative
level.  But another part of the problem are more conceptual issues relating to
understanding quantum theory for a closed system, and furthermore a system that
is not simply sitting in a fixed background spacetime.  These problems have not
yet been satisfactorily solved, even by superstring theory, and they tend to be
the focus of those of us working in quantum cosmology.

Because in certain limits superstring theory reduces to Einstein's general
relativity as an approximation to it, and because many of the more conceptual
issues appear to remain, and indeed stand out more clearly, when one makes the
approximation to Einsteinian gravity without requiring this gravity to be
classical, it is often convenient in quantum cosmology to make the
approximation of taking a quantum version of Einsteinian gravity, quantum
general relativity.  This is not a renormalizable or finite theory, so one
would run into trouble using it in multi-loop calculations that allow
arbitrarily high energy fluctuations, but fortunately many of the more
conceptual issues show up at a cruder approximation, often even at the WKB or
`semiclassical' level with a superposition of fairly classical geometries
(though usually not at the complete semiclassical reduction to a single
classical geometry that solves the classical Einstein equations with a unique
expectation value of the stress-energy tensor as its source).

One standard method of quantizing a spatially closed universe in the
approximation of Einsteinian gravity is canonical quantization.  Since there
exist reviews of this procedure \cite{Hal91,Pag91}, I shall do nothing more
here than to give a very brief sketch of the procedure.  One starts with the
Hamiltonian formulation \cite{ADM}, foliating spacetime into a temporal
sequence of closed spatial hypersurfaces with lapse and shift vectors
connecting them.  Varying the action with respect to these Lagrange multipliers
give the momentum and Hamiltonian constraint equations.  Then one follows the
Dirac method of quantization \cite{Dir}, converting these constraints into
operators and requiring that they annihilate the quantum state.  When the state
is written as a wavefunctional of the three-metric on the spatial
hypersurfaces, the momentum constraint (linear in the momenta) for each point
of space implies that the wavefunctional is unchanged under any series of
infinitesimal diffeomorphisms or coordinate transformations of the three-space
\cite{Mis,Hig}.  The Hamiltonian constraint (quadratic in the momenta) gives
the Wheeler-DeWitt equation \cite{DeW,Whe}, actually one equation for each
point of space, or one equation for each arbitrary choice of the lapse
function.

For each choice of the lapse function, one natural choice for the factoring
ordering
[7,9-13] of the corresponding Wheeler-DeWitt equation turns it into a
Klein-Gordon equation on an indefinite DeWitt metric \cite{DeW} in the
superspace (space of three-metrics), with a potential term that is given by a
spatial integral (weighted by the lapse function) of the spatial curvature plus
the spatial gradient and potential energy terms that go into the energy density
of any matter included in the model.  The WKB approximation for this equation
gives the Hamilton-Jacobi equation for general relativity, and the trajectories
that are orthogonal to the surfaces of constant phase represent classical
spacetimes that solve the classical Einstein equations.

In addition to constraint equations (the Wheeler-DeWitt equations in
canonically quantized general relativity) that restrict the quantum states, a
complete theory of quantum cosmology should specify which solution of these
equations describes our universe.  There have been various proposals for this
in recent years, most notably the Hartle-Hawking `no-boundary' proposal
[14-16] that the wavefunctional evaluated for a compact three-geometry argument
is given by a path integral over compact `Euclidean' four-geometries (with
positive-definite metric signatures, as opposed to `Lorentzian' geometries with
an indefinite signature) which reduce to the compact three-geometry argument at
its boundary, and the Vilenkin tunneling proposal \cite{Vil}, that the
wavefunctional is ingoing into superspace at its regular boundaries and
outgoing at its singular boundaries.

Once one has a quantum state or wavefunctional that satisfies the constraints
of canonical quantum general relativity, there is the question of how to
interpret it to give probabilities.  The first stage of this is to find an
inner product.  DeWitt \cite{DeW} and many others have proposed using the
Klein-Gordon inner product, using the flux of the conserved Klein-Gordon
current.  However, this is not positive definite and vanishes for real
wavefunctionals, such as what one would get from the Hartle-Hawking
`no-boundary' proposal
[14-16].  Another approach is to quantize the true physical variables
\cite{Bar}, which gives unitary quantum gravity, at least at the one-loop
level.  However, it is not clear whether this this approach can be carried
beyond the perturbative level, even in a minisuperspace approximation.  A third
approach is third quantization, in which one converts the Wheeler-DeWitt
wavefunctional to a field operator on superspace.  However, I have not seen any
clear way to get testable probabilities out of this.

I myself favor Hawking's approach \cite{Haw84,HP} of using the na\"{\i}ve inner
product obtained by taking the integral of the absolute square of a
wavefunctional over superspace, with its volume element obtained from the
DeWitt metric on it.  At least this obviously gives a positive-definite inner
product.  Unfortunately, it will almost certainly diverge when the square of a
physical wavefunctional (i.e., one satisfying the Wheeler-DeWitt equations) is
integrated over all of the infinite volume of superspace (which includes
directions corresponding to spatial diffeomorphisms and also to time
translations).  One can try to eliminate the divergences due to the noncompact
diffeomorphism group by integrating only over distinct three-geometries (the
older meaning of superspace) rather than over all three-metrics that overcount
these, though then there is not such an obvious candidate for the volume
element.  However, one would still have divergences due to the noncompactness
of time, which in the canonical quantum gravity of closed universes is encoded
in the three-geometry and matter field configuration on the spatial
hypersurface.  Perhaps one can avoid these divergences by evaluating the inner
product only for wavefunctionals that have been acted on by operators, such as
projection operators, that do not commute with the constraints and which result
in wavefunctionals that are normalizable with the na\"{\i}ve inner product.

\section{Basics of Sensible Quantum Cosmology}

\hspace{.25in}Before going further with trying to calculate probabilities in
quantum cosmology, it may be helpful to ask what probabilities mean in quantum
mechanics.  If they mean propensities for potentialities to be converted to
actualities, then quantum mechanics would seem to be incomplete, since it does
not predict {\it which} possibilities will be actualized, but only the
probabilities for each.  An interpretation in which quantum mechanics would be
more nearly complete would be the Everett or ``many-worlds'' interpretation
\cite{E}, in which all possibilities with positive probabilities are
actualized, though with measures that depend on the quantum probabilities.
However, even this interpretation only seems to work with a single unspecified
set of possibilities, whose probabilities add up to one.  The arbitrariness of
this set seems to indicate that the Everett interpretation is also incomplete,
even though it is more nearly complete than the propensity interpretation.

Therefore, I have proposed a version of quantum mechanics, which I call
Sensible Quantum Mechanics (SQM)
[20-24], in which only a definite set of possibilities have measures, namely,
conscious perceptions.

Sensible Quantum Mechanics is given by the following three fundamental
postulates \cite{P95}:

 {\bf Quantum World Axiom}:  The unconscious ``quantum world'' $Q$ is
completely described by an appropriate algebra of operators and by a suitable
state $\sigma$ (a positive linear functional of the operators) giving the
expectation value $\langle O \rangle \equiv \sigma[O]$ of each operator $O$.

 {\bf Conscious World Axiom}:  The ``conscious world'' $M$, the set of all
perceptions $p$, has a fundamental measure $\mu(S)$ for each subset $S$ of $M$.

 {\bf Quantum-Consciousness Connection}:  The measure $\mu(S)$ for each set $S$
of conscious perceptions is given by the expectation value of a corresponding
``awareness operator'' $A(S)$, a positive-operator-valued (POV) measure
\cite{Dav}, in the state $\sigma$ of the quantum world:
 \begin{equation}
 \mu(S) = \langle A(S) \rangle \equiv \sigma[A(S)].
 \label{eq:1}
 \end{equation}

Here a perception $p$ is the entirety of a single conscious experience, all
that one is consciously aware of or consciously experiencing at one moment, the
total ``raw feel'' that one has at one time, or \cite{Lo} a ``phenomenal
perspective'' or ``maximal experience.''

Since all sets $S$ of perceptions with $\mu(S) > 0$ really occur in SQM, it is
completely deterministic if the quantum state and the $A(S)$ are determined:
there are no random or truly probabilistic elements.  Nevertheless, because SQM
has measures for sets of perceptions, one can readily calculate ratios that can
be interpreted as conditional probabilities.  For example, one can consider the
set of perceptions $S_1$ in which there is a conscious awareness of
cosmological data and theory and a conscious belief that the visible universe
is fairly accurately described by a Friedman-Robertson-Walker (FRW) model, and
the set $S_2$ in which there is a conscious memory and interpretation of
getting reliable data indicating that the Hubble constant is greater than 75
km/sec/Mpc.  Then one can interpret
 \begin{equation}
 P(S_2|S_1)\equiv \mu(S_1\cap S_2)/\mu(S_1)
 \label{eq:2}
 \end{equation}
as the conditional probability that the perception is in the set $S_2$, given
that it is in the set $S_1$, that is, that a perception included a conscious
memory of measuring the Hubble constant to be greater than 75 km/sec/Mpc, given
that one is aware of knowledge and belief that an FRW model is accurate.

\section{Testing Sensible Quantum Cosmology Theories}

\hspace{.25in}The measures for observed (or, perhaps more accurately,
experienced) perceptions can be used to test different SQM theories for the
universe, thus grounding physics and cosmology in experience.    If one had a
theory in which only a small subset of the set of all possible perceptions is
predicted to occur, one could simply check whether an experienced perception is
in that subset.  If it is not, that would be clear evidence against that
theory.  Unfortunately, in almost all SQM theories, almost all sets of
perceptions are predicted to have a positive measure, so these theories cannot
be excluded so simply.  For such many-perceptions theories, the best one can
hope for seems to be to find {\it likelihood} evidence for or against it.  Even
how to do this is not immediately obvious, since SQM theories merely give
measures for sets of perceptions rather than the existence probabilities for
any perceptions (unless the existence probabilities are considered to be unity
for all existing sets of perceptions, i.e., all those with nonzero measure, but
this is of little help, since almost all sets exist in this sense).

In order to test and compare SQM theories, it helps to hypothesize that the set
$M$ of all possible conscious
perceptions $p$ is a suitable topological space with a prior measure
 \begin{equation}
 \mu_0(S) = \int_{S}{d\mu_0(p)}.
 \label{eq:4}
 \end{equation}
Then, because of the linearity of positive-valued-operator measures over sets,
one can write each awareness operator as
 \begin{equation}
 A(S) = \int_S E(p)d\mu_0(p),
 \label{eq:5}
 \end{equation}
a generalized sum or integral of ``experience operators'' or ``perception
operators'' $E(p)$ for the individual perceptions $p$ in the set $S$.
Similarly, one can write the measure on a set of perceptions $S$ as
 \begin{equation}
 \mu(S) =  \langle A(S) \rangle = \int_S d\mu(p)
   = \int_S m(p) d\mu_0(p),
 \label{eq:6}
 \end{equation}
in terms of a measure density $m(p)$ that is the quantum expectation value of
the experience operator $E(p)$ for the same perception $p$:
 \begin{equation}
 m(p) = \langle E(p) \rangle \equiv \sigma[E(p)].
 \label{eq:7}
 \end{equation}

Now one can test the agreement of a particular SQM theory with a conscious
observation or perception $p$ by calculating the (ordinary) {\it typicality}
$T(p)$ that the theory assigns to the perception:  Let $S_{\leq}(p)$ be the set
of perceptions $p'$ with $m(p') \leq m(p)$.  Then
 \begin{equation}
 T(p) \equiv \mu(S_{\leq}(p))/\mu(M).
 \label{eq:13}
 \end{equation}
For $p$ fixed and $\tilde{p}$ chosen randomly with the infinitesimal measure
$d\mu(\tilde{p})$, the probability that $T(\tilde{p})$ is less than or equal to
$T(p)$ is
 \begin{equation}
 P_T(p) \equiv P(T(\tilde{p})\leq T(p)) = T(p).
 \label{eq:14}
 \end{equation}
In the case in which $m(p)$ varies continuously in such a way that $T(p)$ also
varies continuously, this typicality $T(p)$ has a uniform probability
distribution between 0 and 1, but if there is a nonzero measure of perceptions
with the same value of $m(p)$, then $T(p)$ has discrete jumps.  (In the extreme
case in which $m(p)$ has one constant value over all perceptions, $T(p)$ is
unity for each $p$.)

Thus the typicality $T_i(p)$ of a perception $p$ is the probability in a
particular SQM theory or hypothesis $H_i$ that another random perception will
have its measure density and hence typicality less than or equal to that of $p$
itself.  One can interpret it as the likelihood of the perception $p$ in the
particular theory $H_i$, not for $p$ to exist, which is usually unity
(interpreting all perceptions $p$ with $m(p)>0$ as existing), but for $p$ to
have a typicality no larger than it has.

Once the typicality $T_i(p)$ can be calculated for an experienced perception
assuming the theory $H_i$, one approach is to use it to rule out or falsify the
theory if the resulting typicality is too low.  Another approach is to assign
prior probabilities $P(H_i)$ to different theories (presumably neither
propensities nor frequencies but rather purely subjective probabilities,
perhaps one's guess for the ``propensities'' for God to create a universe
according to the various theories), say
 \begin{equation}
 P(H_i)=2^{-n_i},
 \label{eq:16}
 \end{equation}
where $n_i$ is the rank of $H_i$ in order of increasing complexity (my present
favorite choice for a countably infinite set of hypotheses if I could do this
ranking, which is another problem I will not further consider here).  Then one
can use Bayes' rule to calculate the posterior probability of the theory $H_i$
given the perception $p$ as
 \begin{equation}
 P(H_i|p)=\frac{P(H_i)T_i(p)}{\sum_{j}^{}{P(H_j)T_j(p)}}.
 \label{eq:15}
 \end{equation}

There is the potential technical problem that one might assign nonzero
prior probabilities to hypotheses $H_i$ in which the total measure $\mu(M)$ for
all perceptions is {\it not} finite, so that the right side of
Eq.~(\ref{eq:13}) may have both numerator and denominator infinite, which makes
the typicality $T_i(p)$ inherently ambiguous.  To avoid this problem, one might
use, instead of $T_i(p)$ in Eq.~(\ref{eq:15}), rather
 \begin{equation}
 T_i(p;S) = \mu_i(S_{\leq}(p)\cap S)/\mu_i(S)
 \label{eq:17}
 \end{equation}
for some set of perceptions $S$ containing $p$ that has $\mu_i(S)$ finite for
each hypothesis $H_i$.  This is related to a practical limitation anyway, since
one could presumably only hope to be able to compare the measure densities
$m(p)$ for some small set of perceptions rather similar to one's own, though it
is not clear in quantum cosmological theories that allow an infinite amount of
inflation how to get a finite measure even for a small set of perceptions.
Unfortunately, even if one can get a finite measure by suitably restricting the
set $S$, this makes the resulting $P(H_i|p;S)$ depend on this chosen $S$ as
well as on the other postulated quantities such as $P(H_i)$.

Instead of using the particular typicality defined by Eq.~(\ref{eq:13}) above,
one could of course instead use any other property of perceptions which places
them into an ordered set to define a corresponding ``typicality.''  For
example, I might be tempted to order them according to their complexity, if
that could be well defined.  Thinking about this alternative ``typicality''
leaves me surprised that my own present perception seems to be highly
complicated but apparently not infinitely so.  What simple complete theory
could make a typical perception have a high but not infinite complexity?

However, the ``typicality'' defined by Eq.~(\ref{eq:13}) has the merit of
being defined purely from the prior and fundamental measures, with no added
concepts such as complexity that would need to be defined.  The necessity of
being able to rank perceptions, say by their measure density, in order to
calculate a typicality, is indeed one of my main motivations for postulating a
prior measure given by Eq.~(\ref{eq:4}).

Nevertheless, there are alternative typicalities that one can define purely
from the prior and fundamental measures.  For example, one might define a {\it
reversed typicality} $T_r(p)$ in the following way (again assuming that the
total measure $\mu(M)$ for all perceptions is finite):  Let $S_{\geq}(p)$ be
the set of perceptions $p'$ with $m(p') \geq m(p)$.  Then
 \begin{equation}
 T_r(p) \equiv \mu(S_{\geq}(p))/\mu(M).
 \label{eq:13r}
 \end{equation}
For $p$ fixed and $\tilde{p}$ chosen randomly with the infinitesimal measure
$d\mu(\tilde{p})$, the probability that $T_r(\tilde{p})$ is less than $T_r(p)$
is
 \begin{equation}
 P_{T_r}(p) \equiv P(T_r(\tilde{p})\leq T_r(p)) = T_r(p),
 \label{eq:14r}
 \end{equation}
the analogue of Eq.~(\ref{eq:14}) for the ordinary typicality.

In the generic continuum case in which $m(p)$ varies continuously in such a
way that there is only an infinitesimally small measure of perceptions whose
$m(p)$ are infinitesimally near any fixed value, the reversed typicality
$T_r(p)$ is simply one minus the ordinary typicality, i.e., $1-T(p)$, and also
has a uniform probability distribution between 0 and 1.  Its use arises from
the fact that just as a perception with very low ordinary typicality $T(p)\ll
1$ could be considered unusual, so a perception with an ordinary typicality too
near one (and hence a reversed typicality too near zero, $T_r(p)\ll 1$) could
also be considered unusual, ``too good to be true.''

Perhaps one might like to combine the ordinary typicality with the reversed
typicality to say that a perception giving either typicality too near zero
would be evidence against the theory.  For example, one might define the {\it
dual typicality} $T_d(p)$ as the probability that a random perception
$\tilde{p}$ has the lesser of its ordinary and its reversed typicalities less
than or equal to that of the perception under consideration:
 \begin{equation}
 T_d(p) \equiv P(\min{[T(\tilde{p}),T_r(\tilde{p})]}
 \leq   \min{[T(p),T_r(p)}])
 \equiv \mu(S_d(p))/\mu(M),
 \label{eq:14d}
 \end{equation}
where $S_d(p)$ is the set of all perceptions $\tilde{p}$ with the minimum of
its ordinary and reversed typicalities less than or equal to that of the
perception $p$, i.e., the set with
$\min{[T(\tilde{p}),T_r(\tilde{p})]}\leq\min{[T(p),T_r(p)]}$.  In the case in
which $T(p)$, and hence also $T_r(p)$, varies continuously from 0 to 1,
 \begin{equation}
 T_d(p) = 1 - | 1 - 2T(p) |.
 \label{eq:14e}
 \end{equation}
Then the dual typicality $T_d(p)$ would be very small if the ordinary
typicality $T(p)$ were very near either 0 or 1.

Of course, one could go on with an indefinitely long sequence of typicalities,
say making a perception ``atypical'' if $T(p)$ were very near any number of
particular values at or between the endpoint values 0 and 1.  But these
endpoint values are the only ones that seem especially relevant, and so it
would seem rather {\it ad hoc} to define ``typicalities'' based on any other
values.  Since $T_d(p)$ is symmetrically defined in terms of both endpoints
(or, more precisely, in terms of both the $\leq$ and the $\geq$ relations for
$m(p')$ in comparison with $m(p)$), in some sense it seems the most natural one
to use.  Obviously, one could use it, or its modification along the lines of
Eq.~(\ref{eq:17}), instead of $T(p)$ in the Bayesian Eq.~(\ref{eq:15}).

To illustrate how one may use these typicalities to test different theories,
consider the experiment in which one makes a particular measurement of a single
particular continuous variable (e.g., the position of a one-dimensional
harmonic oscillator in its ground state) which is supposed to have a gaussian
quantum distribution.  Suppose that in the subset of perceptions $S$ in which
this measurement is believed to have been made and the result is known, there
is a one-dimensional continuum of perceptions that are linearly related to the
measured value of the variable, say labeled by the real number $x$ that is the
measured value of the variable, with the prior measure $d\mu_0(p)\propto dx$.
(It might be more realistic to suppose that the measuring device can transmit
only a discrete set of values to the brain that is doing the perceiving, but if
the number of these is large, it is convenient to approximate them by a
continuum.  As to whether the set of perceptions is discrete or continuous, I
know of no strong evidence either way.  Even if the possible quantum states of
the universe lay in a finite-dimensional Hilbert space, which seems doubtful
but is not clearly ruled out by observational evidence, one could easily have a
continuum of experience operators $E(p)$ for a continuum set of perceptions
$p$.)

I shall furthermore make the idealized hypothesis that the measure for this
subset of perceptions labeled by $x$ is purely given by the quantum
distribution of the measured variable and is not further influenced by
processes in the brain.  One could imagine situations in which certain measured
values lead to the release of an anaesthetic that renders the perceiver
unconscious and so essentially eliminates the measure for the corresponding
perceptions.  At first sight such situations that grossly affect the measure
for perceptions arising from a quantum measurement seemed contrived, but then I
realized that noticeably unusual results of the measuring device (say results
several standard deviations from the mean) could very well attract more
conscious attention, over a longer time, than results that are not noticeably
unusual.  It seems highly plausible that the measure for a set of perceptions
would increase with their alertness and with the time over which they
continually occur, so noticeably unusual events that attract more attention
would presumably have a higher measure than one might otherwise na\"{\i}vely
expect.  Thus the measure for perceptions arising from the measurement of a
variable with a gaussian quantum distribution could well have tails that do not
decrease so fast as the original gaussian.

I shall call this effect, whereby the measure for perceptions of unusual events
is increased by the attention given them, the Attention Effect.  It may well be
responsible for the large number of coincidences that one is aware of from
anecdotal evidence.  (For example, it has occurred to me several times that it
was surprising for Canada and the U.S. to have ages in years that were both
perfect cubes, 125 and 216 respectively, in 1992.)  One can try to combat it,
by, e.g., focusing on perceptions in which it is perceived that the quantum
measurement was made only a second previous, say, when there would presumably
not have been time for a dull result to have receded much from consciousness.
So for simplicity in the following discussion I shall make the idealized
assumption that the Attention Effect, as well as any other effect that distorts
the measure of perceptions from what what one would calculate from the quantum
measurement itself, is negligible.  (I am grateful for the visit of Jane and
Tim Hawking in Edmonton during my writing of the previous paragraph, which gave
me the time to realize the importance of the Attention Effect.  Was the timing
of their arrival another coincidence?)

Assuming no Attention Effect or similar effect, the measure for $x$ would have
the gaussian distribution it inherits from the quantum measurement, say
 \begin{equation}
 m(x) \propto e^{-x^2/(2\sigma^2)}.
 \label{eq:55}
 \end{equation}
Within the subset of perceptions $S$ in which this measurement is believed to
have been made and the result is known, the ordinary, reversed, and dual
typicalities are
 \begin{equation}
 T(x;S) ={\rm erfc}(\sqrt{x^2/(2\sigma^2)})\equiv
 1-{\rm erf}(\sqrt{x^2/(2\sigma^2)})\equiv
 1-{2 \over \sqrt{\pi}}
 \int_{0}^{\sqrt{x^2/(2\sigma^2)}}{e^{-z^2}dz},
 \label{eq:14f}
 \end{equation}
 \begin{equation}
 T_r(x;S) =1-T(x;S)={\rm erf}(\sqrt{x^2/(2\sigma^2)}),
 \label{eq:14g}
 \end{equation}
 \begin{eqnarray}
 T_d(x;S) = 1 - | 1 - 2T(x;S) | = 1 - | 1 - 2T_r(x;S) |
 \nonumber \\=1 - | 1 - 2 {\rm erfc} (\sqrt{x^2/(2\sigma^2)}) |
 = 1 - | 1 - 2 {\rm erf} (\sqrt{x^2/(2\sigma^2)}) |.
 \label{eq:14h}
 \end{eqnarray}

Suppose that one does not know the actual standard deviation $\sigma$ for $x$
but has various hypotheses $H_i$ that it has the various values $\sigma_i$.
Then one can replace $\sigma$ with $\sigma_i$ in
Eqs.~(\ref{eq:14f})-(\ref{eq:14h}) above to get the typicalities (restricted to
the subset $S$) of the perception $p$ that gives the perception component $x$
in the corresponding hypothesis.  If the typicality one is considering is too
low for a certain hypothesis, one might use the perception to exclude (or
falsify on a likelihood basis) that hypothesis.

Alternatively, one might assign a prior probability distribution to $\sigma_i$
and then use the Bayesian Eq.~(\ref{eq:15}), with $T_i(p)$ replaced by
$T(x;S)$, $T_r(x;S)$, or $T_d(x;S)$ as given by
Eqs.~(\ref{eq:14f})-(\ref{eq:14h}) and with $\sigma$ replaced by $\sigma_i$, to
calculate a posterior distribution for $\sigma_i$, given $x$ from the
perception.  For example, one might assign (as a fairly simple concrete choice)
the following prior probability distribution for $\sigma_i$, which is purely a
decaying exponential distribution in its square and thus a slight modification
of a gaussian distribution for $\sigma_i$ itself:
 \begin{equation}
 P(\sigma_i) =
 e^{-\sigma_i^2/(2\sigma_0^2)}
 \sigma_i d\sigma_i/\sigma_0^2.
 \label{eq:56}
 \end{equation}
Here $\sigma_0^2$ is an arbitrary parameter that one must choose to represent
the expectation value of $\sigma_i^2$ in the prior distribution.  One can then
readily calculate \cite{GR} that using the ordinary typicality in
Eq.~(\ref{eq:15}) gives
 \begin{equation}
 P(\sigma_i|x;S) =
 e^{(x/\sigma_0)-\sigma_i^2/(2\sigma_0^2)}
 {\rm erfc}(\sqrt{x^2/(2\sigma^2)})
 \sigma_i d\sigma_i/\sigma_0^2.
 \label{eq:57}
 \end{equation}

This is very strongly damped for small values of $\sigma_i$ (i.e., for values
much less than $x$), by the complementary error function from
Eq.~(\ref{eq:14f}) for the typicality, and is damped at large values of
$\sigma_i$ (i.e., for values much greater than $\sigma_0$) by the exponentially
decaying prior distribution of Eq.~(\ref{eq:56}).  However, if one used a
different prior distribution that was not significantly damped at large values
of $\sigma_i$, neither would the posterior distribution be significantly damped
at large values of $\sigma_i$.  Thus using the ordinary typicality $T$ in the
Bayesian Eq.~(\ref{eq:15}) is effective in giving a lower limit on the spread
of the measure distribution for a number like $x$ assigned to the perceptions
(at least if the one used in the calculation is not exactly at the mean value),
but it is not effective in giving a better upper limit on the spread than that
given by the prior distribution.  This is because the ordinary typicality gives
a penalty for theories that would predict that an observed result is unlikely
or ``bad,'' but it does not give a similar penalty for theories that predict
that the observed result is ``too good to be likely.''

The reversed typicality $T_r$ of Eq.~(\ref{eq:13r}) or (\ref{eq:14g}) does
indeed penalize theories that predict that the observed result is too good to
be likely, but only at the cost of not penalizing at all ``bad'' results, so it
would be worse to use.  Better is the dual typicality $T_d$ of
Eq.~(\ref{eq:14d}) or (\ref{eq:14h}), which penalizes theories that fit the
observations either too poorly or too well.  Unfortunately, it makes it harder
to evaluate the posterior probability distribution Eq.~(\ref{eq:15})
analytically, because of the minimization functions in the definition of the
dual typicality, and I have been unable to come up with a simple explicit
result for the prior distribution Eq.~(\ref{eq:56}), though one can get an
explicit result for a prior distribution that is flat in $n\equiv 1/\sigma_i^2$
\cite{P95}.  In any case, one can see that with the dual typicality, the
posterior distribution for the standard deviation $\sigma_i$ is more heavily
damped at both large and small $\sigma_i$ than is the prior distribution for
$\sigma_i$ that is assumed.

One can see that if one starts with a smooth continuum prior probability
distribution for theories, a Bayesian analysis using one of the typicalities of
an experienced perception can give a posterior probability distribution of
theories that is more narrow, but it can never lead to a nonzero probability
for any single theory out of the continuum of possibilities that are smoothly
weighted.  If this is the case, one shall never succeed in getting any single
final theory that one can say has any significant (e.g., nonzero) probability
of being absolutely correct.  Of course, one might be able to deduce (after
postulating a reasonable continuum prior distribution) fairly narrow ranges for
the continuous parameters where most of the posterior probability is
concentrated, so the situation could be qualitatively no worse than it is at
present for such parameters as the fine structure constant, which is only
thought to be likely to be within some very narrow range (and perhaps only
within this range for certain components of the quantum state of the universe
from which our perceptions get the bulk of their measure).

On the other hand, it is conceivable that theorists will eventually find a
discrete set of theories, each with no arbitrary continuous parameters (as in
superstring theories), or else with preferred discrete values of such
parameters, to each of which they can assign a nonzero prior probability.  Then
if these are weighted by the likelihoods they predict for the perception of a
sufficiently good set of observations, it is conceivable (though at present it
might seem somewhat miraculous, but doesn't the order and structure in our
world already look miraculous?) that the posterior probabilities will pick out
one unique theory with a probability near unity and assign all other theories a
total probability much closer to zero.  In such a case theorists might well
believe that the one unique theory with a probability near unity is indeed {\it
the correct theory} of the universe.  Of course, the fact that the other
theories would not have a total probability exactly equal to zero (at least for
almost any conceivable scenario I can presently imagine) would mean that one
could not be sure (at least by the Bayesian analysis outlined above) that one
really did have the correct theory for the universe, but if the probability
were sufficiently near unity, one could presumably put a great deal of faith in
that deduction from theory and observation, just as we presently put a great
deal of faith in much smaller pieces of knowledge to which we assign
probabilities near unity.

There is also the problem that the prior probabilities seem to be purely
subjective, so people could well disagree about whether or not the assignment
that led to a posterior probability near unity for one particular theory (if
indeed that dream of my present wishful thinking actually does occur) is
reasonable.  I suspect that such disagreements about prior probabilities are at
the heart of many current disagreements (e.g., the existence of God or the
truth of superstring theory), though there is also the huge practical problem
that unless one has a detailed theory from which one can make the relevant
calculations, one cannot even predict what the likelihoods are for the various
hypotheses to result in certain observations (e.g., the perception of good and
evil or the experience of gravity).  However, one could imagine a society which
agrees in sufficient broad outline about the prior probabilities, and
observational results that sufficiently narrow them down (as often occurs to a
fantastic degree in experimental physics, viz. the amazing agreement of
experiment with QED), that nearly all members will agree on a unique theory for
the universe as having a posterior probability near unity (similar to the fact
that most members of the community of physicists believes in QED within its
domain of applicability, though QED represents a continuum of theories labeled
by the fine structure constant rather than one unique theory).

One can also worry that if Sensible Quantum Mechanics is correct, then
presumably our perceptions are completely determined by the detailed theory,
and it seems likely that it would only be some sort of an idealization to say
that our beliefs are determined by a Bayesian process of modifying prior
probabilities to posterior probabilities in the light of the likelihoods of our
observations.  So it may be purely hypothetical to predict what beliefs we
would arrive at if we went though this ideal Bayesian procedure, since it would
seem miraculous if we were determined to act in just that way, and there is
certainly evidence that most of the time our beliefs are not determined
precisely thus.  However, idealizations are at the heart of physics, and this
Bayesian one does not seem particularly worse than many others.  Even if the
physics community (or the broader human community) does not actually come to
its beliefs by precisely a Bayesian analysis, it is interesting to speculate on
what conclusions it might eventually arrive at if it did.  Though even such
conclusions are not guaranteed to be true (because of the difficulties
mentioned above, and various others), it would seem that they would be more
likely to be true than whatever conclusions people will actually come up with,
especially if they they choose to ignore the Bayesian procedure.

Thus, although Sensible Quantum Mechanics is my current best guess of what is a
true framework for a complete description of the universe, my advocacy of a
Bayesian analysis to choose between detailed theories within that framework is
not a guess of what is truly the way physicists work, but a moral appeal for
one way in which I think the search for a detailed theory ought to be
conducted.

\section{Predictions from Sensible Quantum Cosmology}

\hspace{.25in}Although it is certainly an open question whether humans or any
other conscious beings within the universe will ever come to some sort of a
grasp of a complete theory of the universe (perhaps only as a set of ideas
whose logical implications include a complete description of the entire
universe, even though it seems extremely unlikely that conscious beings within
the universe could ever work out all of these implications in detail), we would
like to develop better theories that will enable us to predict more properties
of the universe than we can at present.

Vilenkin \cite{Vil95} has recently discussed predictions in quantum cosmology,
and I do not have much in detail to add to what he beautifully covered.  I
agree with him that if the `constants' of physics that we have measured can
actually vary from component to component of the quantum state of the universe,
the relevant probability distribution must be obtained by using something like
the Principle of Mediocrity that he proposes, that we are a `typical'
civilization within the ensemble of components.

I might personally prefer \cite{P95,P95c} a slightly different variant of the
Weak Anthropic Principle
[29-35] which I call the Conditional Aesthemic Principle, that our conscious
perceptions are likely to be typical perceptions in the conscious world with
its measure.  Then the relevant probability distribution for the `constants' of
physics would be weighted by the measures for perceptions.  The most basic way
to do this would be to use only the perceptions which include a belief in the
value of the corresponding `constant.'  However, if the quantum state of the
universe is represented by the density matrix $\rho$, and if the `constants'
are indeed constant over the relative density
matrix \cite{P95,P95c}
 \begin{equation}
 \rho_p=\frac{E(p)\rho E(p)}{Tr[E(p)\rho E(p)]}
 \label{eq:P4}
 \end{equation}
for each perception $p$, then it might be better to weight the probability
distribution of the `constants' in each such relative state by the measure
density for the corresponding perception.  One might like thus to find out the
probability distribution for the cosmological constant, the parameters of the
Standard Model of particle physics, and the parameters of an inflaton potential
(if any such exists).

One persistent problem that hampers predictions even from simple toy
minisuperspace models in quantum cosmology is the apparent lack of
normalizability of most of the quantum states, at least if one takes the
positive-definite na\"{\i}ve inner product obtained by integrating the absolute
square of the wavefunctional over superspace.  As discussed above, part of this
problem in canonical quantum cosmology is due to the fact that the wavefunction
in the configuration representation is most easily handled as a functional of
three-metrics that is invariant under coordinate transformations, but this
diffeomorphism group is noncompact and leads to divergences when one integrates
over it.   It seems that this ought to be merely an avoidable technical problem
that arises from writing down wavefunctionals of three-metrics rather than of
three-geometries, but even if one circumvents it (or avoids it by truncating
the configuration space, as in homogeneous minisuperspace models), one next
faces the problem of the invariance under the transformations generated by the
Wheeler-DeWitt equations, which represent time translations at each point of
space.  It would seem that then one must face at least the noncompact groups of
Lorentz boosts at each point of space, since the hypersurface described by the
three-metric argument of the wavefunctional can be tilted rather arbitrarily at
each point of space.  Even if those divergences are eliminated (as in the
minisuperspace models in which the homogeneous three-geometries are prevented
from being tilted), one can still get divergences from the infinite range of
values allowed by the `time.'

Perhaps these divergences can be avoided by not seeking to interpret the
integral of the absolute square of the physical wavefunctional over the
configuration space, but by restricting the interpretation, as in Sensible
Quantum Mechanics, to the integrals of wavefunctionals that have been operated
on by the experience or perception operators $E(p)$.  There seems to be no
reason why the resulting wavefunctionals should be physical wavefunctionals
that obey the constraints such as the Wheeler-DeWitt equation, since a
perception could very well be localized to be concentrated near a particular
clock time (represented by some property of the three-geometry or matter fields
on it) and need not occur over the whole history (or, more accurately, set of
histories in the path integral sense) of the universe that is represented by a
physical solution of the constraints.  Certainly in minisuperspace one can get
positive operators, as the experience operators are required to be, that have
finite expectation values even for wavefunctionals that are not normalizable,
e.g., projection operators to finite ranges of all the configuration space
variables.

However, I am still not sure that even this procedure will avoid all
divergences.  If one has a model in which an infinite amount of inflation can
occur, which would be useful for solving the flatness problem, then in the
resulting infinite amount of spatial volume it would seem that the measure for
almost any set of perceptions of nonzero prior measure would likely be
infinite, since each sufficiently large finite volume would seem likely to give
an independent positive contribution to it.  If one has infinite measures for
almost all nontrivial sets of perceptions, then calculating conditional
probabilities and typicalities by taking the ratios of such infinite quantities
is meaningless, and I do not yet see how to get testable predictions out from
even Sensible Quantum Cosmology.

In conclusion, Sensible Quantum Mechanics seems to give an improved
interpretation of quantum theory that is helpful in quantum cosmology for
testing different quantum theories of the universe and in making predictions.
It seems to ameliorate part of the problem of the lack of normalizability of
the wavefunctionals of canonical quantum gravity, but serious problems still
seem to remain with this that hamper predictions from quantum cosmology.

People whom I remember to have recently influenced my thoughts on this subject
are listed in \cite{P95}.  I am grateful to Alex Vilenkin for sending me an
advance copy of the third paper of \cite{Vil95}.  Financial support has been
provided by the Natural Sciences and Engineering Research Council of Canada.



\baselineskip 7pt

\begin{thebibliography}{99}

\bibitem{Hal91} J. J. Halliwell, in {\em Quantum Cosmology and Baby Universes},
edited by S. Coleman, J. B. Hartle, T. Piran, and S. Weinberg (World
Scientific, Singapore, 1991), p. 159.

\bibitem{Pag91} D. N. Page, in {\em Gravitation:  A Banff Summer Institute},
edited by R. Mann and P. Wesson (World Scientific, Singapore, 1991), p. 135.

\bibitem{ADM} R. Arnowitt, S. Deser, and C. W. Misner, in {\em Gravitation:  An
Introduction to Current Research}, edited by L. Witten (Wiley, New York, 1962).

\bibitem{Dir} P. A. M. Dirac, Proc. Roy. Soc. (Lond.) {\bf A246}, 326  and 333
(1958); {\em Lectures on Quantum Mechanics} (Yeshiva University, New York,
1964).

\bibitem{Mis} C. W. Misner, Rev. Mod. Phys. {\bf 29}, 497 (1957).

\bibitem{Hig} P. W. Higgs, Phys. Rev. Lett. {\bf 1}, 373 (1959).

\bibitem{DeW} B. S. DeWitt, Phys. Rev. {\bf 160}, 1113 (1967).

\bibitem{Whe} J. A. Wheeler, in {\em Battelle Rencontres:  1967 Lectures in
Mathematics and Physics}, edited by C. DeWitt and J. A. Wheeler (Benjamin, New
York, 1968).

\bibitem{Mis72} C. W. Misner, in {\em Magic without Magic}, edited by J. R.
Klauder (Freeman, San Francisco, 1972).

\bibitem{Kuc} K. Kucha\v{r}, in {\em Relativity, Astrophysics and Cosmology},
edited by W. Israel (Reidel, Dordrecht, 1973).

\bibitem{HPT} M. Henneaux, M. Pilati, and C. Teitelboim, Phys. Lett. {\bf
110B}, 123 (1982).

\bibitem{CZ} T. Christodoulakis and J. Zanelli, Phys. Rev. D{\bf 29}, 2738
(1984); Phys. Lett. {\bf 102A}, 227 (1984).

\bibitem{HP} S. W. Hawking and D. N. Page, Nucl. Phys. {\bf B264}, 185 (1986).

\bibitem{Haw82} S. W. Hawking, in {\em Astrophysical cosmology:  Proceedings of
the Study Week on Cosmology and Fundamental Physics} edited by H. A. Br\"{u}ck,
G. V. Coyne, and M. S. Longair (Pontificiae Academiae Scientiarum Scripta
Varia, Vatican, 1982).

\bibitem{HH} J. B. Hartle and S. W. Hawking, Phys. Rev. D{\bf 28}, 2960 (1983).

\bibitem{Haw84} S. W. Hawking, Nucl. Phys. {\bf B239}, 257 (1984).

\bibitem{Vil} A. Vilenkin, Phys. Lett. {\bf 117B}, 25 (1982); Phys. Rev. D{\bf
30}, 509 (1984); Nucl. Phys. {\bf B252}, 141 (1985); Phys. Rev. D{\bf 33}, 3560
(1986); Phys. Rev. D{\bf 37}, 888 (1988).

\bibitem{Bar} A. O. Barvinsky, in {\em Proceedings of the Fourth Seminar on
Quantum Gravity, May 25-29, 1987, Moscow, USSR}, edited by M. A. Markov, V. A.
Berezin, and V. P. Frolov (World Scientific, Singapore, 1988); Nucl. Phys. {\bf
B325}, 705 (1989); Phys. Lett. {\bf 241B}, 201 (1990); Phys. Rep. {\bf 230},
237 (1993).

\bibitem{E} H. Everett, III, Rev.\ Mod.\ Phys. {\bf 29}, 454 (1957); B. S.
DeWitt and N. Graham, eds., {\em The Many-Worlds Interpretation of Quantum
Mechanics} (Princeton University Press, Princeton, 1973).

\bibitem{P94a} D. N. Page, ``Probabilities Don't Matter,'' to be published in
{\em Proceedings of the 7th Marcel Grossmann Meeting on General Relativity},
edited by M. Keiser and R. T. Jantzen (World Scientific, Singapore 1995)
(University of Alberta report Alberta-Thy-28-94, Nov. 25, 1994), gr-qc/9411004.

\bibitem{P94b} D. N. Page, ``Information Loss in Black Holes and/or Conscious
Beings?'' to be published
in {\em Heat Kernel Techniques and Quantum Gravity}, edited by S. A. Fulling
(Discourses in Mathematics and Its Applications, No. 4, Texas A\&M University
Department of Mathematics, College Station, Texas, 1995) (University of Alberta
report Alberta-Thy-36-94, Nov. 25, 1994), hep-th/9411193.

\bibitem{P95} D. N. Page, ``Sensible Quantum Mechanics:  Are Only Perceptions
Probabilistic?'' (University of Alberta
report Alberta-Thy-05-95, June 7, 1995), quant-ph/9506010.

\bibitem{P95b} D. N. Page, ``Attaching Theories of Consciousness to Bohmian
Quantum Mechanics,'' to be published in {\em Bohmian Quantum Mechanics and
Quantum Theory:  An Appraisal}, edited by J. T. Cushing, A. Fine, and S.
Goldstein (Kluwer, Dordrecht, 1996) (University of Alberta report
Alberta-Thy-12-95, June 30, 1995), quant-ph/9507006.

\bibitem{P95c} D. N. Page, ``Sensible Quantum Mechanics:  Are Probabilities
Only in the Mind?'' to be published in {\em Proceedings of the Sixth Seminar on
Quantum Gravity, June 12-19, 1987, Moscow, Russia}, edited by V. A. Berezin and
V. A. Rubakov (World Scientific, Singapore, 1996) (University of Alberta report
Alberta-Thy-13-95, July 4, 1995), gr-qc/9507024.

\bibitem{Dav} E. B. Davies, {\em Quantum Theory of Open Systems} (Academic
Press, London, 1976).

\bibitem{Lo} M. Lockwood, {\em Mind, Brain and the Quantum:  The Compound `I'}
(Basil Blackwell, Oxford, 1989).

\bibitem{GR} I. S. Gradshteyn and I. M. Ryzhik, {\em Table of Integrals,
Series, and Products, Corrected and Enlarged Edition}, edited by A. Jeffrey
(Academic Press, San Diego, 1980), item 6.284, p. 649.

\bibitem{Vil95} A. Vilenkin, Phys. Rev. Lett. {\bf 74}, 846 (1995); ``Making
Predictions in Eternally Inflating Universe,'' gr-qc/9505031; ``Predictions
from Quantum Cosmology'' (Lectures at International School of Astrophysics `D.
Chalonge,' Erice, 1995).

\bibitem{D} R. H. Dicke, Rev.\ Mod.\ Phys.\ {\bf 29}, 355 and 363 (1977);
Nature {\bf 192} 440 (1961).

\bibitem{Ca} B. Carter, in {\em Confrontation of Cosmological Theories with
Observation}, edited by M. S. Longair (Reidel, Dordrecht, 1974), p.~291.

\bibitem{CR} B. J. Carr and M. J. Rees, Nature {\bf 278}, 605 (1979).

\bibitem{Ro} I. L. Rozental, Sov.\ Phys.\ Usp.\ {\bf 23}, 296 (1980).

\bibitem{Da} P. C. W. Davies, {\em The Accidental Universe} Cambridge
University Press, Cambridge, 1982).

\bibitem{BT} J. D. Barrow and F. T. Tipler, {\em the Anthropic Cosmological
Principle} (Clarendon Press, Oxford, 1986).

\bibitem{Les} J. Leslie, Am.\ Phil.\ Quart., 141 (April 1982); Mind, 573
(October 1983); in {\em Current Issues in Teleology}, edited by N. Rescher
(University Press of America, Lanham and London, 1983), p.~111; in {\em
Proceedings of the Philosophy of Science Association 1986} (Edwards Bros, Ann
Arbor, 1986), vol.~1, p.~87; in {\em Origin and Early History of the Universe},
edited by J. Demaret (University of Li\`{e}ge, Li\`{e}ge, 1987), p.~439; Mind,
269 (April 1988); {\em Universes} (Routledge, London and New York, 1989); {\em
Physical Cosmology and Philosophy} (Macmillan, New York, 1990).

\end{thebibliography}
\end{document}